# Clustering Students According to their Academic Achievement Using Fuzzy Logic

**Serhiy Balovsyak**
Yuriy Fedkovych Chernivtsi National University, Chernivtsi, 58012, Ukraine
E-mail: s.balovsyak@chnu.edu.ua
ORCID iD: https://orcid.org/0000-0002-3253-9006

**Oleksandr Derevyanchuk\***
Yuriy Fedkovych Chernivtsi National University, Chernivtsi, 58012, Ukraine
E-mail: o.v.derevyanchuk@chnu.edu.ua
ORCID iD: https://orcid.org/0000-0002-3749-9998
*Corresponding Author

**Hanna Kravchenko**
High State Educational Establishment «Chernivtsi transport college», Chernivtsi, 58000, Ukraine
E-mail: hannakravchenko81@gmail.com
ORCID iD: https://orcid.org/0009-0004-7609-0345

**Yuriy Ushenko**
Yuriy Fedkovych Chernivtsi National University, Chernivtsi, 58012, Ukraine
E-mail: y.ushenko@chnu.edu.ua
ORCID iD: https://orcid.org/0000-0003-1767-1882

**Zhengbing Hu**
School of Computer Science, Hubei University of Technology, Wuhan, China
E-mail: drzbhu@gmail.com
ORCID iD: https://orcid.org/0000-0002-6140-3351

**Abstract:** The software for clustering students according to their educational achievements using fuzzy logic was developed in Python using the Google Colab cloud service. In the process of analyzing educational data, the problems of Data Mining are solved, since only some characteristics of the educational process are obtained from a large sample of data. Data clustering was performed using the classic K-Means method, which is characterized by simplicity and high speed. Cluster analysis was performed in the space of two features using the machine learning library scikit-learn (Python). The obtained clusters are described by fuzzy triangular membership functions, which allowed to correctly determine the membership of each student to a certain cluster. Creation of fuzzy membership functions is done using the scikit-fuzzy library. The development of fuzzy functions of objects belonging to clusters is also useful for educational purposes, as it allows a better understanding of the principles of using fuzzy logic. As a result of processing test educational data using the developed software, correct results were obtained. It is shown that the use of fuzzy membership functions makes it possible to correctly determine the belonging of students to certain clusters, even if such clusters are not clearly separated. Due to this, it is possible to more accurately determine the recommended level of difficulty of tasks for each student, depending on his previous evaluations.

## 1. Introduction

At present, large volumes of data are processed in education, which are obtained at various stages of the educational process [1-4]. For example, the parameters of students and the results of their studies in various subjects are

                    

processed. When analyzing such data, the problems of Data Mining are solved, since only some useful characteristics need to be obtained from a large sample of data [5-7]. Manual analysis of educational data is quite time time-consuming and requires a high qualification of the performer. Therefore, computer clustering methods are currently widely used for the analysis of educational data.

Clustering methods use unsupervised learning, so no prior training is required [8,9]. Clustering methods are used to divide the initial set of large data into parts (clusters). This allows you to structure the data and identify certain useful patterns in them. In addition, the analysis of a small cluster is much simpler than the processing of the entire initial set. Clustering is usually performed in the space of two parameters. As clustering parameters, for example, intermediate and final results of students' studies in certain subjects are used. A number of clustering methods are used. A popular and effective clustering method is the K-means. This method allows to divide the initial set of data into several clusters that do not overlap, that is, clear clustering is performed. As a result of clustering students according to their grades, each cluster contains students with a characteristic combination of grades, for example, with average grades for the first assignment and high grades for the second assignment.

In an ideal case, the inter-cluster distance is much larger than the intra-cluster distance, so such clusters are clearly separated. However, when processing real educational data, there are often no clear boundaries between clusters. The problem is that objects on the borders of clusters are difficult to attribute to only one cluster. Such data clusters can be considered as fuzzy sets, for which the membership of an object to a certain cluster is described by a fuzzy membership function. Due to fuzzy membership functions, it is possible to correctly describe the membership of a certain object, which is on the border of clusters, to several clusters at the same time. Therefore, the purpose of this work is to increase the accuracy of clustering of educational data due to the use of fuzzy logic tools [10].

Thus, the work is relevant because the use of fuzzy membership functions allows to correctly determine the membership of students to certain clusters, even if such clusters are not clearly separated. Thanks to this, it is possible to more accurately determine the recommended level of difficulty of tasks for each student, depending on his educational achievements.

## 2. Related Works

Clustering methods are effective for the analysis of educational data, as they allow dividing a set of objects into groups (clusters) according to certain parameters. A characteristic feature of clusters is that objects in one cluster are more similar to each other than objects in different clusters. Cluster analysis of data in order to evaluate and increase the efficiency of the educational process was carried out in a number of works [8,9,10,11,12].

Currently, a number of clustering methods are used, each of which has certain advantages, disadvantages and scope of use. For example, the machine learning library scikit-learn (in Python) contains more than 10 clustering methods. Source [8] describes the main methods of clustering:

1. The classic K-Means clustering method is widespread, simple and fast. For these reasons, it is advisable to use the K-Means method for clustering educational data. In this method, the number of clusters is set by the user based on the conditions of the problem or by sorting in a certain range. The division of objects into clusters is performed iteratively. The obtained clusters do not intersect, that is, one object can belong to only one cluster. For each cluster, the coordinates of its center (centroid) are calculated.

2. The MeanShift method is similar to K-Means, but determines the number of clusters automatically. A common field of use of the method is segmentation.

3. The DBSCAN (Density-Based Application Spatial Clustering with Noise) method is based on the assumption that clusters are dense regions in the feature space that are separated by less dense regions.

4. The BIRCH (Balanced Iterative Reducing and Clustering using Hierarchies) method belongs to the hierarchical method, it is quite effective from the point of view of memory usage when processing large data sets. The BIRCH method creates a tree-like data structure and calculates the coordinates of the centroids.

In work [12], clustering methods were used to analyze student behavioral patterns. Clustering methods of 5 categories are considered: partitioning methods, density-based methods, grid-based methods and model-based methods, hierarchical methods. It is shown that different clustering methods can give different results when processing the same set of data. Hierarchical clustering is considered separately, which consists in building a hierarchy of clusters using two methods: agglomerative (based on merging) and divisive (based on splitting data). It is shown that combining the results of various clustering methods, for example, Density-Based Spatial Clustering of Applications with Noise (DBSCAN) and k-means, is effective.

In modern educational platforms, clustering methods (for example, the K-Means method) provide an effective method of evaluating student learning outcomes [9]. Based on the analysis of such clustering, recommendations for training are provided both in the case of face-to-face and distance learning. Cluster analysis of intermediate and final learning results allows adjusting the level of difficulty of learning tasks. The application of intelligent data analysis to educational statistics in general expands the possibilities of network education, increases the effectiveness of teaching and learning. In the process of intelligent data analysis, hidden, unknown and potentially useful information is obtained from a large database.

In the field of education, clustering is used to identify student behavior patterns and take appropriate actions to optimize the learning process. Cluster analysis makes it possible to identify various behavioral factors that are strongly correlated with the success of students' studies [10]. In most cases, K-Means, BIRCH, and DBSCAN algorithms are used for clustering educational data [11]. With their help, it is possible to analyze and correct students' studies in electronic systems, as well as predict their achievements.

In order to increase the speed of data processing, parallel clustering methods are used, for example, the DBSCAN parallel method [13]. The greatest advantage of the DBSCAN method is robustness to outliers.

However, when clustering educational data, the obtained clusters are not always clearly separated, and the parameters of the educational process are often vague values [14]. In this case, the criteria of the education system are described as fuzzy sets with certain fuzzy membership functions (for example, triangular). In work [15], an analysis of the success of students in higher education was implemented using fuzzy logic. Student success in this work is described by fuzzy membership functions.

Fuzzy clustering algorithms (soft clustering algorithms), such as Fuzzy K-Means and Fuzzy C-Means (FCM) algorithms, are used to process data that do not form clear clusters [16]. The Fuzzy K-Means algorithm is a fuzzy analogue of the K-Means algorithm. A feature of the Fuzzy C-Means algorithm is the automatic determination of the number of clusters. Fuzzy clustering algorithms allow one element of the set to belong to several clusters (with different degrees of belonging) [17]. Fuzzy logic capabilities are widely used to control complex systems [18]. In work [19] the use of fuzzy membership functions for image segments is considered. However, membership functions calculated by fuzzy clustering algorithms do not in all cases correctly describe the membership of a certain object to different clusters, taking into account the requirements of a specific task. This is caused, in particular, by the complex and asymmetric (relative to the center) form of clusters obtained during the processing of educational data. Therefore, to expand the possibilities of analysis, the description of data by means of fuzzy logic is used, with the possibility of choosing the parameters of fuzzy membership functions.

Based on the analysis of modern research, a conclusion was made about the need to develop software tools for clustering students based on the results of their studies using fuzzy logic. It is advisable to perform clustering using the classic K-Means method, which belongs to the model-based category, is characterized by simplicity and high speed. The obtained clusters should be described by fuzzy membership functions, which will allow to correctly determine the membership of each student to a certain cluster. The development of fuzzy membership functions of objects belonging to clusters is also useful for educational purposes, as it allows a better understanding of the principles of using fuzzy logic. Software tools for processing educational data should be developed in Python using specialized libraries and modules (for example, scikit-learn library, module sklearn. cluster).

## 3. Methodology

### 3.1. Basics of educational data clustering

The educational process and its participants will be described using $D$ parameters $y_1, y_2,..., y_D$, each of which takes $Q$ values. For example, the parameters $y_1$ and $y_2$ could represent the $Q$ scores of students in Math and English, respectively. For the sake of simplification, the cluster analysis will be carried out in the space of two features $x$ and $y$, that is, we will simultaneously carry out clustering according to 2 parameters (for example, for $x = y_1$, $y = y_2$) [8]. Thus, if perform clustering according to $x$ and $y$ parameters, then each object of educational data can be represented as a point on a plane with coordinates $(x, y)$.

Let us consider in more detail the clustering according to the parameters $x(i)$ and $y(i)$, where $i = 1,..., Q$. To find the distance between objects with numbers $i$ and $j$, will use the Euclidean distance $\rho(i, j)$:

$$\rho(i,j) = \sqrt{(x(i) - x(j))^2 + (y(i) - y(j))^2}. \tag{1}$$

Intra-cluster distance (2) and inter-cluster distance (3) are used to assess the quality of clustering. The average intra-cluster distance $F_0$ should be as small as possible:

$$F_0 = \frac{\sum_{i=1}^{Q}\sum_{j=1}^{Q}[i=j]\rho(i,j)}{\sum_{i=1}^{Q}\sum_{j=1}^{Q}[i=j]} \rightarrow min. \tag{2}$$

The average inter-cluster distance $F_1$ should be as large as possible:

$$F_1 = \frac{\sum_{i=1}^{Q}\sum_{j=1}^{Q}[i \neq j]\rho(i,j)}{\sum_{i=1}^{Q}\sum_{j=1}^{Q}[i \neq j]} \rightarrow max. \tag{3}$$

In practice, the ratio of a pair of functionals is calculated to take into account both inter-cluster $F_1$ and intra-cluster distances $F_0$:

$$F_0/F_1 \to min. \tag{4}$$

The minimization of relation (4) is implemented programmatically in the K-Means function of the scikit-learn library [20].

### 3.2. Basics of using fuzzy membership functions

Fuzzy triangular membership functions were used to determine the belonging of a certain object to clusters [15,16]. The fuzzy membership function $\mu_x(k, dx)$ describes the membership to the cluster with number $k$ of the object whose distance to the center of cluster $k$ along the $x$ coordinate is equal to $dx$ (the object number is denoted by $iv$). Similarly, the fuzzy membership function $\mu_y(k, dy)$ describes the membership to the cluster with the number $k$ of the object whose distance to the center of the cluster $k$ along the $y$ coordinate is equal to $dy$. The fuzzy membership function $\mu_\rho(k, \rho)$ describes the membership to the cluster with the number $k$ of the object for which the Euclidean distance (1) to the center of the cluster $k$ is equal to $\rho$.

For each cluster with number $k$, its radius $R_c$ is calculated as the Euclidean distance (1) from the center of the cluster in the space of two features $x$ and $y$. Forms of clusters for real educational data differ in significant asymmetry, therefore, in addition to the radius $R_c$, the following radii were calculated for each cluster:

- $R_{cxL}$ – radius for the left part of the cluster according to the $x$ coordinate.
- $R_{cxR}$ – radius for the right part of the cluster according to the $x$ coordinate.
- $R_{cyDn}$ – radius for the lower part of the cluster according to the $y$ coordinate.
- $R_{cyUp}$ – the radius for the upper part of the cluster according to the $y$ coordinate.

The calculation of fuzzy membership functions of objects belonging to clusters is performed as follows. Based on the Euclidean distance $\rho$ of the object to the center of the cluster $k$ by coordinates $(x, y)$, the degree of belonging of the object to the cluster is calculated as the value of the fuzzy membership function $\mu_\rho(k, \rho)$ according to the formula:

$$\mu_\rho(k, \rho) = 1 - \frac{\rho}{R_c \cdot k_R}, \tag{5}$$

where $R_c$ is the radius of cluster $k$, calculated as the Euclidean distance (1);
$k_R$ – radius change factor (for example, $k_R = 1.5$).

By the radius change factor $k_R$, it is possible to adjust the overlap of clusters depending on the conditions of the task (the stronger the relationship of the data, the more overlapping the clusters and the larger $k_R$ value). That is, if the distance $\rho$ of the object from the center of cluster $k$ is equal to 0, then $\mu_\rho(k, \rho) = 1$; if the distance $\rho$ is equal to or greater than $R_c \cdot k_R$, then $\mu_\rho(k, \rho) = 0$ (the object does not belong to the cluster).

Similarly to the function $\mu_\rho(k, \rho)$, the following fuzzy membership functions are calculated:

- $\mu_x(k, dx)$ is a fuzzy membership function of object belonging to cluster $k$, whose distance to the center of the cluster by coordinate $x$ is equal to $dx$.
- $\mu_y(k, dy)$ is a fuzzy membership function of object belonging to cluster $k$, the distance of which to the center of the cluster according to the $y$ coordinate is equal to $dy$.

The calculation of $\mu_x(k, dx)$ for the left part of the cluster (relative to the center) is performed according to the formula:

$$\mu_x(k, dx) = 1 - \frac{dx}{R_{cxL} \cdot k_R}, \tag{6}$$

and for the right part of the cluster is performed according to a similar formula

$$\mu_x(k, dx) = 1 - \frac{dx}{R_{cxR} \cdot k_R}. \tag{7}$$

The calculation of $\mu_y(k, dy)$ for the lower part of the cluster (relative to the center) is performed according to the formula:

$$\mu_y(k, dy) = 1 - \frac{dy}{R_{cyDn} \cdot k_R}, \tag{8}$$

and for the upper part of the cluster according to a similar formula

$$\mu_y(k, dy) = 1 - \frac{dy}{R_{cyUp} \cdot k_R}.$$ (9)

The degree of belonging of an object to cluster $k$ by all coordinates $\mu_{xy}(k, dx, dy)$ is calculated using the values of two membership functions $\mu_x(k, dx)$ and $\mu_y(k, dy)$ according to the formula:

$$\mu_{xy}(k, dx, dy) = \frac{\sqrt{(\mu_x(k,dx))^2 + (\mu_y(k,dy))^2}}{\sqrt{2}}.$$ (10)

By correction of the coefficient $k_R$ in formulas (5-9), it is possible to clarify the values of fuzzy membership functions, to regulate more or less overlap of clusters. Thus, the proposed methodology allows supplementing clustering methods with fuzzy logic, which provides more accurate cluster analysis for educational data.

## 4. Software Implementation

Software tools for the clustering of educational data and their analysis using fuzzy logic were developed in the Python language using the Google Colab cloud service (in the Jupyter Notebook) [21]. Data clustering consists in dividing $Q$ objects of the educational process into $Q_k$ clusters in the space of two features (based on parameters $x$ and $y$).

According to the flowchart of the algorithm (Fig. 1), initial educational data (for example, student grades in various subjects) are first read. Among all the parameters of the educational process, two are chosen: $x$ and $y$. If the range of values of parameter $x$ is significantly larger (more than an order of magnitude) than the range of values of parameter $y$, then one of the parameters is scaled so that the ranges of values of different parameters differ by no more than an order of magnitude (such scaling is required to take into account when clustering the values of both parameters). Next, the user sets the quantity of clusters $Q_k$.

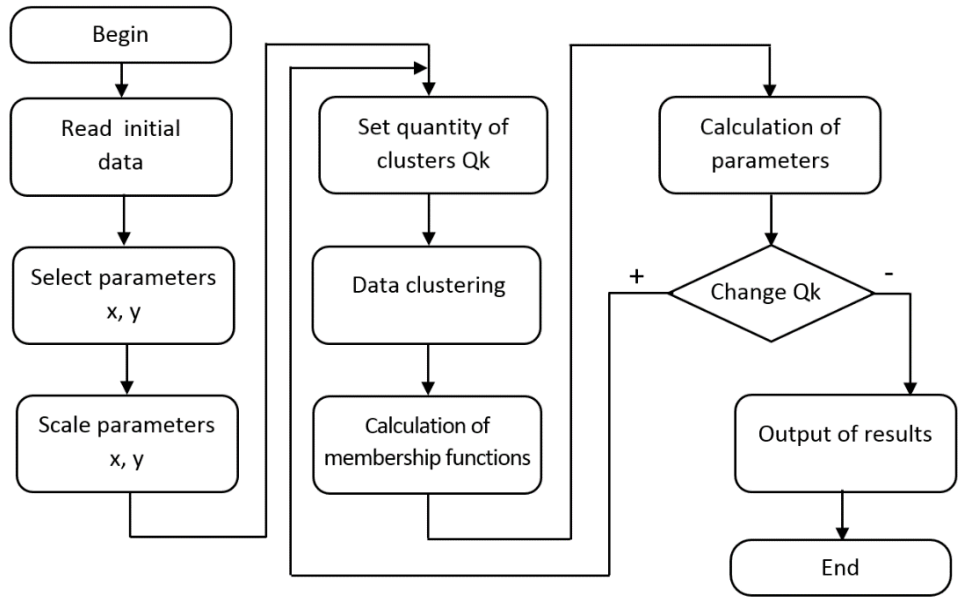

Fig. 1. Flowchart of the algorithm of cluster analysis of educational data

Data clustering (recorded in arrays $x(i)$ and $y(i)$, where $i = 1,..., Q$) is performed by the K-Means method by the KMeans() function of the sklearn.cluster library using the Euclidean distance (1) between objects (in the space of their parameters). In the process of clustering, relation (4) is maximized. The number of clusters $Q_k$ is passed as a parameter to the KMeans() function. As a result of clustering, the coordinates of the centers of the clusters (as the centers of gravity) are calculated: the coordinates of the centers of the clusters according to the first parameter are written into the array $C_s(k, 0)$, and the coordinates of the centers of the clusters according to the second parameter are written into the array $C_s(k, 1)$, where the cluster number $k = 1,..., Q_k$. The cluster number for each object is recorded in the array $N_c(i)$, where $i = 1,..., Q$. The number of objects $C_q$ in the clusters is also calculated.

After clustering, the cluster radii $R_c$, $R_{cxL}$, $R_{cxR}$, $R_{cyDn}$ and $R_{cyUp}$ are calculated. For each cluster $k$, by formulas (5-10) calculated fuzzy membership functions $\mu_p(k, p)$, $\mu_x(k, dx)$, $\mu_y(k, dy)$ and $\mu_{xy}(k, dx, dy)$. Calculation of triangular fuzzy membership functions is performed by the trimf() function of the scikit-fuzzy library. Fuzzy membership functions quantitatively describe the membership of an object to a certain cluster (the smaller the distance from the object to the center of the cluster, the greater the value of the corresponding membership functions). On the basis of fuzzy membership functions, the degree of membership of objects specified by the user to each cluster is determined. The resulting clusters and fuzzy membership functions are visualized.

If the selected number of clusters $Q_k$ does not provide the required accuracy of clustering (taking into account criterion (4)), then clustering is performed for the changed value of $Q_k$. The obtained results of the cluster analysis are stored in files.

## 5. Helpful Hints

### 5.1. Initial data

Verification of the developed software was performed using test educational data (Table 1) [22]. The learning results of each student with number $i$ are described by a number of parameters: "math score", "reading score" and "writing score".

Table 1. A fragment of initial educational data [22]

| Student number $i$ | "Parental level of education" | "Math score" | "Reading score" | "Writing score" |
|---|---|---|---|---|
| 1 | "bachelor's degree" | "72" | "72" | "74" |
| 2 | "some college" | "69" | "90" | "88" |
| 3 | "master's degree" | "90" | "95" | "93" |
| 4 | "associate's degree" | "47" | "57" | "44" |
| 5 | "some college" | "76" | "78" | "75" |
| 6 | "associate's degree" | "71" | "83" | "78" |
| 7 | "some college" | "88" | "95" | "92" |
| 8 | "some college" | "40" | "43" | "39" |
| 9 | "high school" | "64" | "64" | "67" |
| 10 | "high school" | "38" | "60" | "50" |
| 11 | "associate's degree" | "58" | "54" | "52" |
| 12 | "associate's degree" | "40" | "52" | "43" |
| 13 | "high school" | "65" | "81" | "73" |
| 14 | "some college" | "78" | "72" | "70" |
| 15 | "master's degree" | "50" | "53" | "58" |
| 16 | "some high school" | "69" | "75" | "78" |
| 17 | "high school" | "88" | "89" | "86" |
| 18 | "some high school" | "18" | "32" | "28" |
| 19 | "master's degree" | "46" | "42" | "46" |
| 20 | "associate's degree" | "54" | "58" | "61" |
| 21 | "high school" | "66" | "69" | "63" |
| 22 | "some college" | "65" | "75" | "70" |
| 23 | "some college" | "44" | "54" | "53" |
| 24 | "some high school" | "69" | "73" | "73" |
| 25 | "bachelor's degree" | "74" | "71" | "80" |
| 26 | "master's degree" | "73" | "74" | "72" |
| 27 | "some college" | "69" | "54" | "55" |
| 28 | "bachelor's degree" | "67" | "69" | "75" |
| 29 | "high school" | "70" | "70" | "65" |
| 30 | "master's degree" | "62" | "70" | "75" |
| 31 | "some college" | "69" | "74" | "74" |

The cluster analysis was carried out on the basis of the numerical fields named "math score", "reading score" and "writing score", the values of which are in the range from 0 to 100. The value of the field "math score" is read as a parameter $y_1$, the value of the field "reading score" are read as the $y_2$ parameter, and the "writing score" field values are read as the $y_3$ parameter. Each of the parameters $y_1$, $y_2$ and $y_3$ takes $Q$ values ($Q = 1000$).

### 5.2. Results of cluster analysis

A cluster analysis was carried out in the space of two features with the names "math score" and "reading score", that is, the values of the parameters $x = y_1$ and $y = y_2$. Using the K-Means method, the following results of initial data clustering were obtained (Fig. 2).

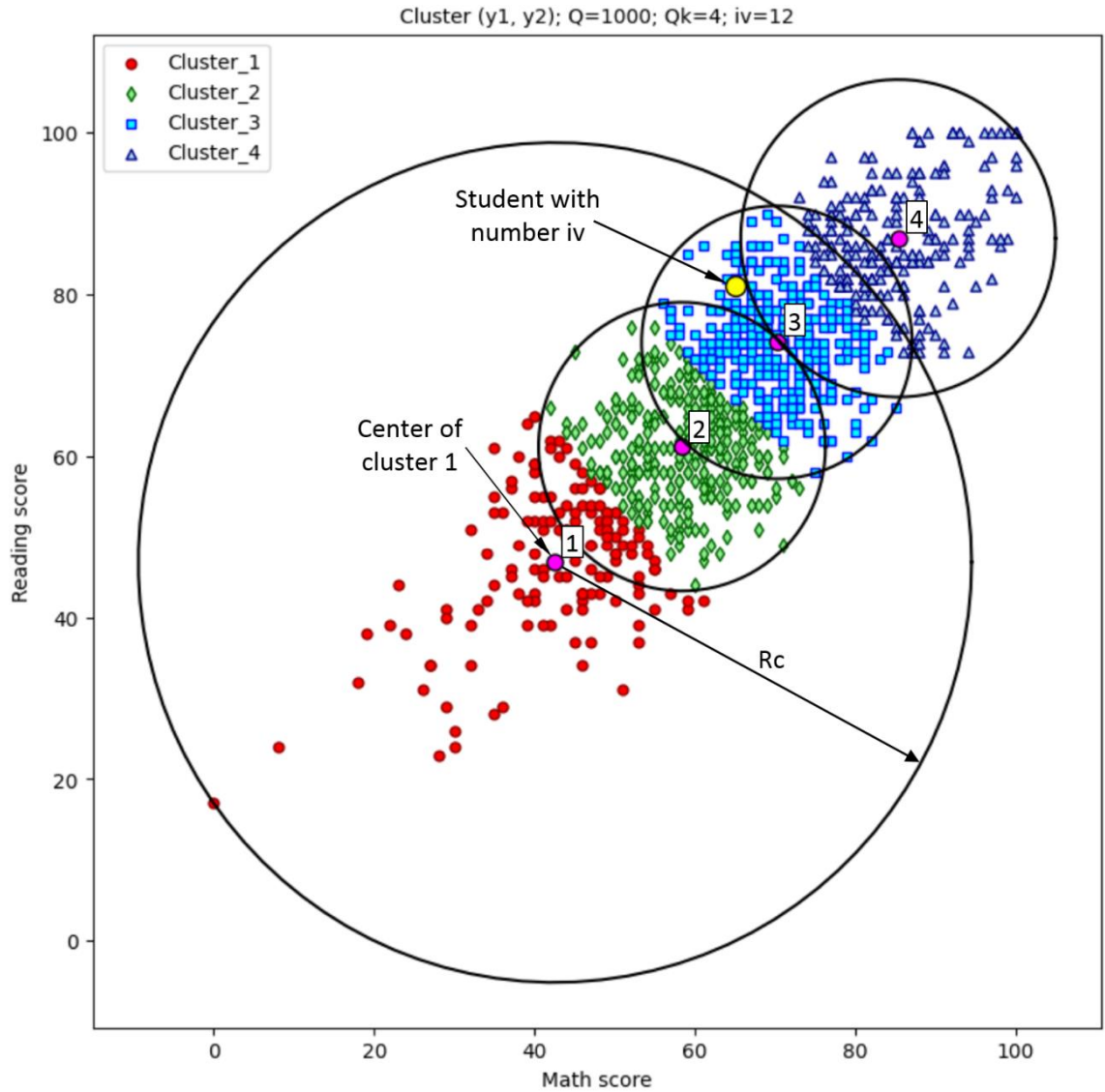

Fig. 2. An example of clustering of test data (Table 1) according to the parameters "math score", "reading score"; Rc is the radius of the cluster

Each number of the cluster with the number $k$ corresponds to a set that describes the results of student learning:

- $k = 1$ – the set of "low math and reading scores".
- $k = 2$ – the set of "below average math and reading scores".
- $k = 3$ – the set of "average math and reading scores".
- $k = 4$ – the set of "high math and reading scores".

Therefore, on the basis of the obtained clusters and their corresponding sets, it is possible to evaluate the results of student learning.

The forms of clusters for real educational data differ in significant asymmetry (Fig. 2), therefore, in addition to the radius $R_c$, the radii $R_{cxL}$, $R_{cxR}$, $R_{cyDn}$ and $R_{cyUp}$ were also calculated for each cluster. The introduction of such 4 radii ($R_{cxL}$, $R_{cxR}$, $R_{cyDn}$, $R_{cyUp}$) makes it possible to more accurately localize the areas of the clusters (Fig. 3).

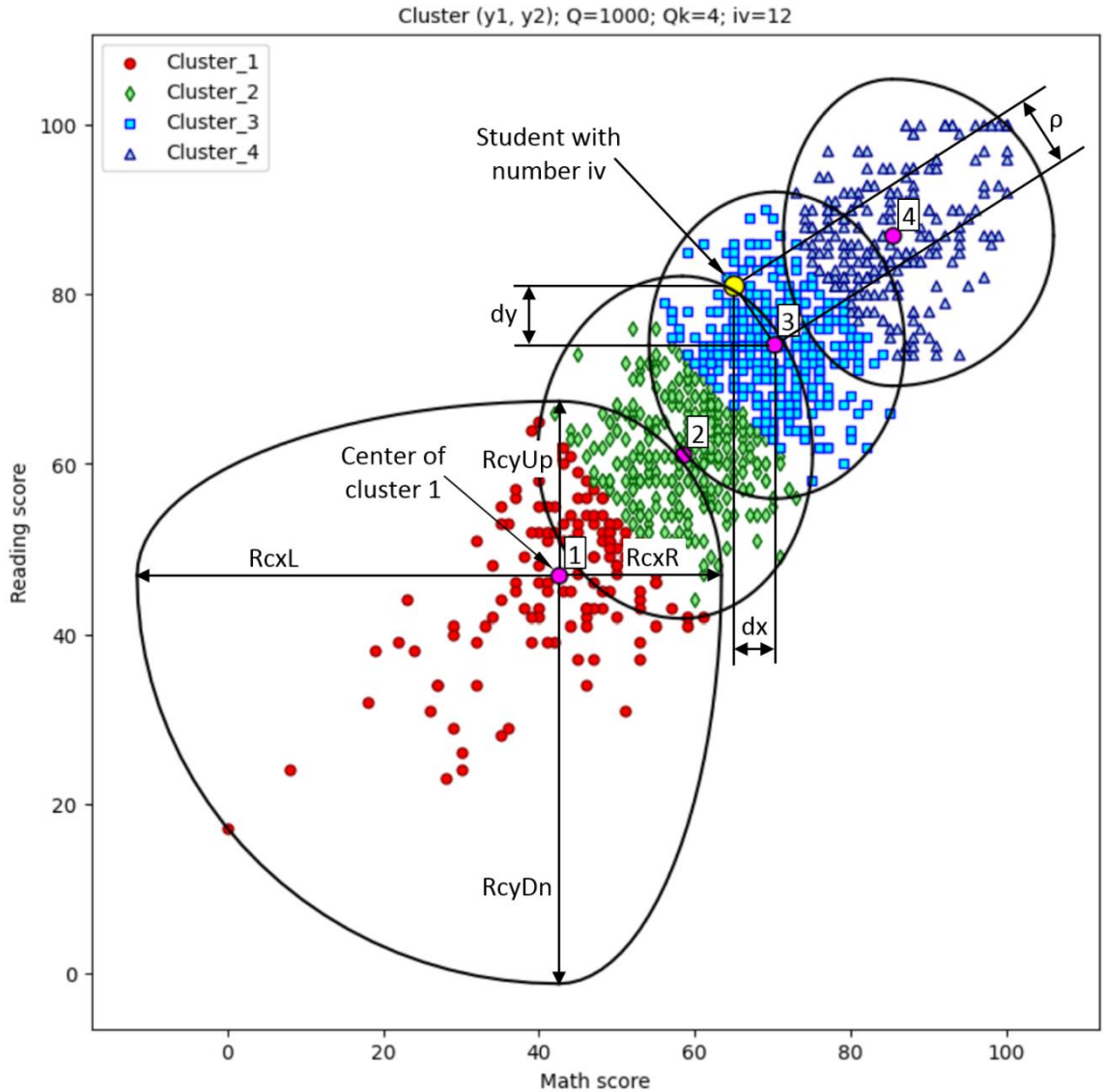

Fig. 3. An example of clustering of test data (Table 1) according to the parameters "math score", "reading score"; cluster boundaries are marked by arcs with radii RcxL, RcxR, RcyDn, RcyUp (in each quadrant of the coordinate system)

The distance from the center of the cluster to an arbitrary object (any cluster) along the *x* coordinate is denoted by *dx* (Fig. 3). Similarly, the distance from the center of the cluster to an arbitrary object along the *y* coordinate is denoted by *dy*, and the Euclidean distance from the center of the cluster to an arbitrary object is denoted by ρ.

For the user-selected student with number $i_v$, the distance $dx0$ to the center of cluster *k* is calculated by coordinate *x*, and based on the distance $dx0$, the degree of belonging of this student to cluster *k* is calculated as the value $\mu_x(k, dx0)$ (Fig. 4). For the selected student with number $i_v$, the distance $dy0$ to the center of cluster *k* along the *y* coordinate is also calculated, and based on the distance $dy0$, the degree of belonging of this student to cluster *k* is calculated as the value $\mu_y(k, dy0)$ (Fig. 5). Similarly, for the selected student with number $i_v$, the Euclidean distance $\rho_0$ from the center of the cluster is calculated, and based on the distance $\rho_0$, the degree of belonging of this student to cluster *k* is calculated as the value $\mu_\rho(k, \rho_0)$ (Fig. 6).

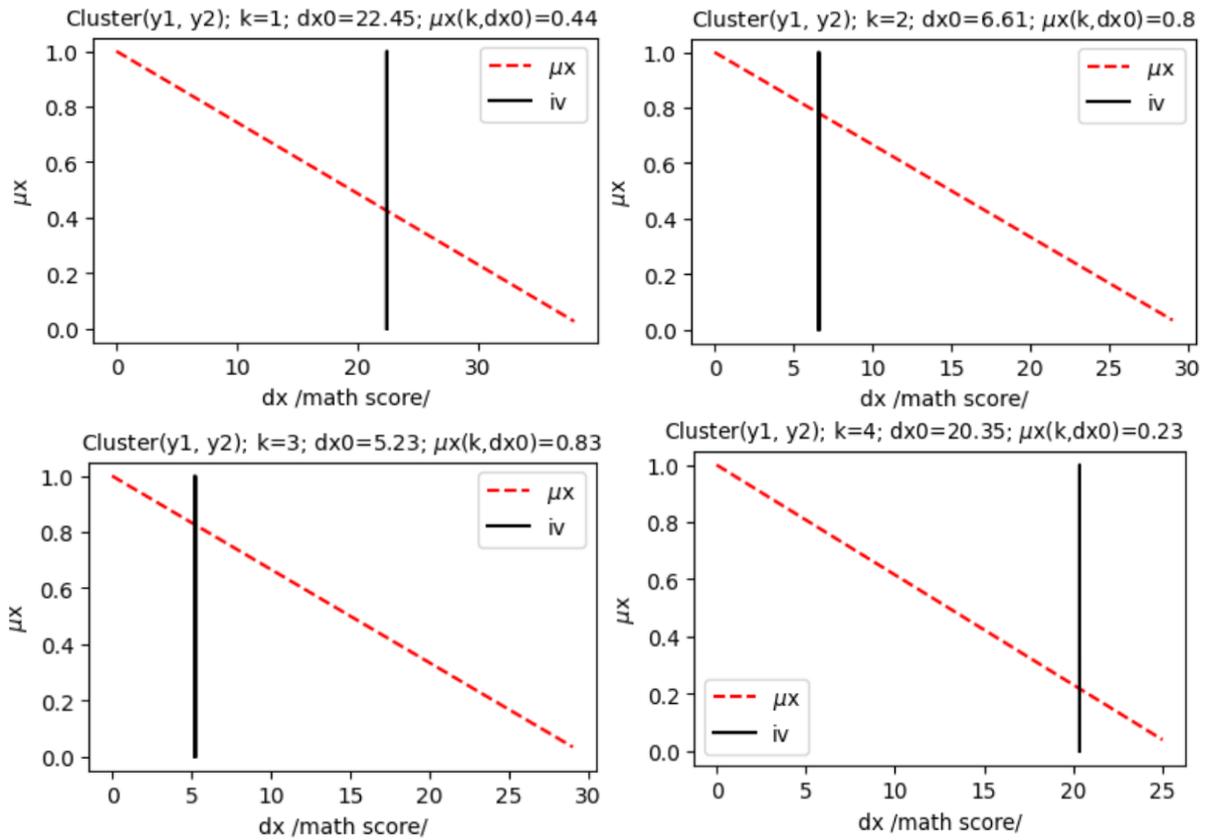

Fig. 4. Fuzzy membership functions $\mu_x(k, dx)$ of students to clusters with numbers $k$ = 1, 2, 3, 4 according to the "math score" parameter; the values of the distance $dx0$ for the student with number $i_v$ =12 are indicated by a vertical line (Fig. 3)

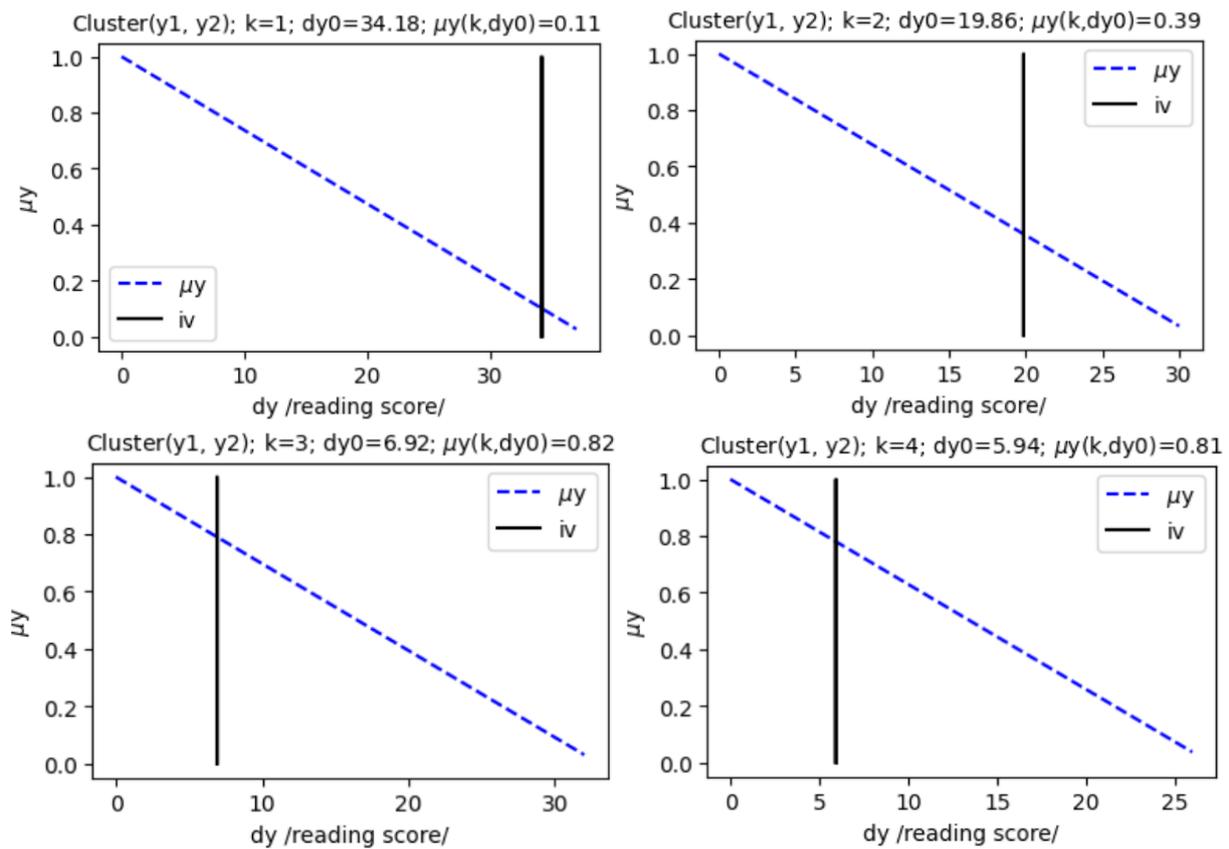

Fig. 5. Fuzzy membership functions $\mu_y(k, dy)$ of students to clusters with numbers $k$ =1, 2, 3, 4 according to the "reading score" parameter; the values of the distance $dy0$ for the student with the number $i_v$ =12 are indicated by a vertical line (Fig. 3)

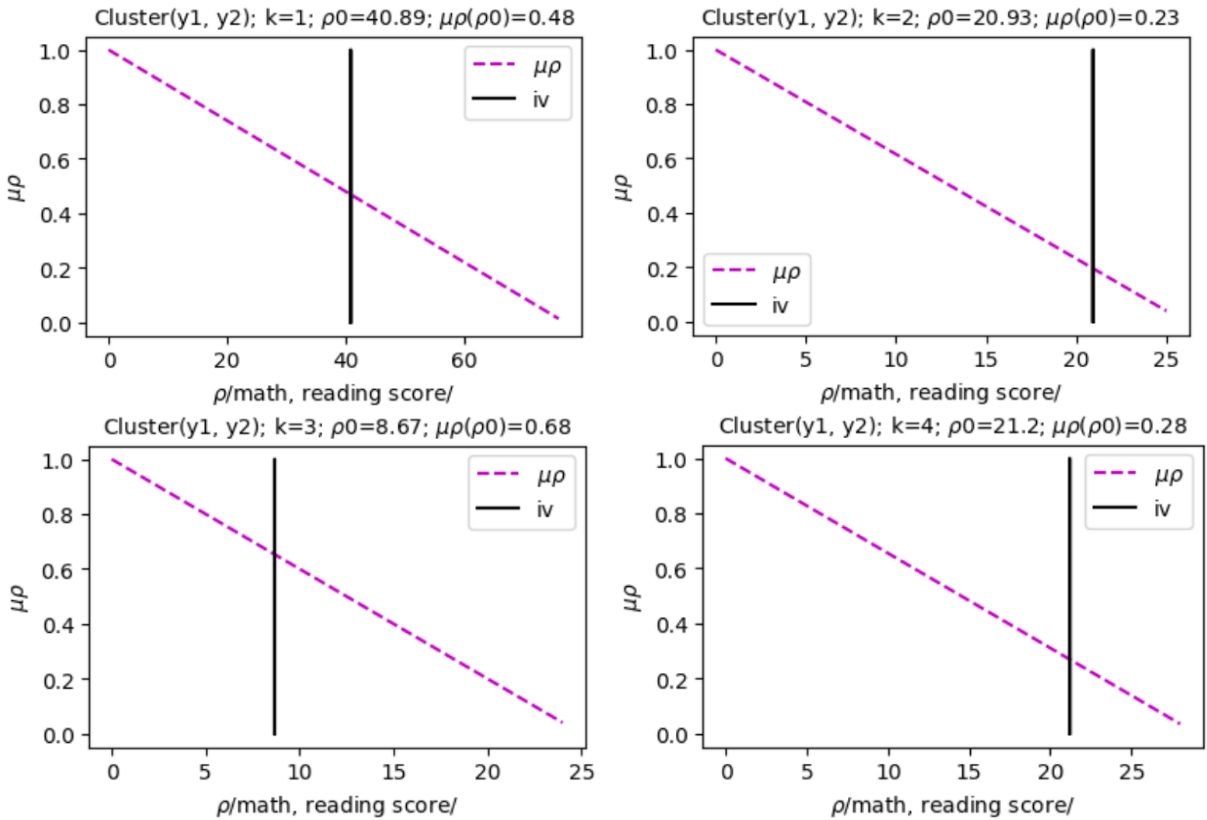

Fig. 6. Fuzzy membership functions $\mu_\rho(k, \rho)$ of students to clusters with numbers $k$ =1, 2, 3, 4 according to "math score" and "reading score" parameters; the values of the distance $\rho0$ for the student with number $i_v$ =12 are indicated by a vertical line (Fig. 3)

Based on the fuzzy membership functions $\mu_x(k, dx0)$ (Fig. 4) and $\mu_y(k, dy0)$ (Fig. 5), the value of the degree of belonging of the student with number $i_v$ (selected by the user) to clusters with numbers $k$ (Fig. 7) was calculated as the values of $\mu_{xR}(k)$ and $\mu_{yR}(k)$, respectively. On the basis of fuzzy membership functions $\mu_\rho(k, \rho0)$ (Fig. 6), the value of the degree of belonging of the student with number $i_v$ to clusters with numbers $k$ (Fig. 8) was calculated as the value of $\mu_{\rho R}(k)$. Based on the calculated values of $\mu_{xR}(k)$ and $\mu_{yR}(k)$ by formula (10), the value of the degree of belonging of the student with number $i_v$ to clusters with numbers $k$ was calculated as the value of $\mu_{xyR}(k)$. At the same time, $\mu_{xyR}(k)$ values more correctly show the belonging of objects to clusters (compared to $\mu_{\rho R}(k)$), since the asymmetric shape of clusters, which is described by the radii $R_{cxL}$, $R_{cxR}$, $R_{cyDn}$, $R_{cyUp}$, is taken into account when calculating $\mu_{xyR}(k)$.

Therefore, based on the calculated $\mu_{xyR}(k)$ values, it is possible to fairly accurately assess the student's learning outcomes. For example, the student with number $i_v$ (Fig. 8) mostly belongs to cluster #3 ($\mu_{xyR}(3) = 0.83$) – the fuzzy set "average math and reading scores", and to a lesser extent belongs to cluster #2 ($\mu_{xyR}(2) = 0.63$) is a fuzzy set of "below average math and reading scores". Such information is also useful for determining the level of difficulty of educational tasks (each number of the cluster corresponds to the corresponding level of difficulty of the task). For example, for the considered student with number $i_v$, it is advisable to offer tasks with difficulty level 3 and additionally tasks with difficulty level 2.

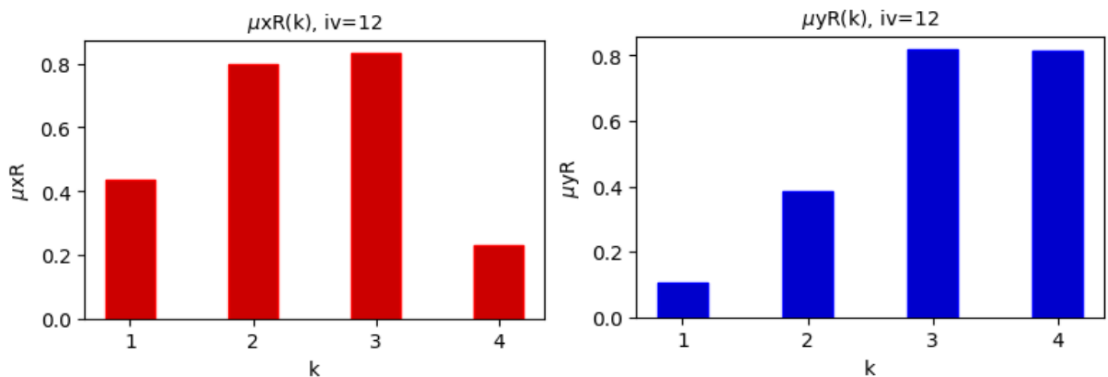

Fig. 7. The value of the degree of belonging of a student with number $i_v$ =12 (Fig. 3) to clusters with numbers $k$, calculated on the basis of fuzzy membership functions $\mu_x(k, dx0)$ (Fig. 4) and $\mu_y(k, dy0)$ (Fig. 5)

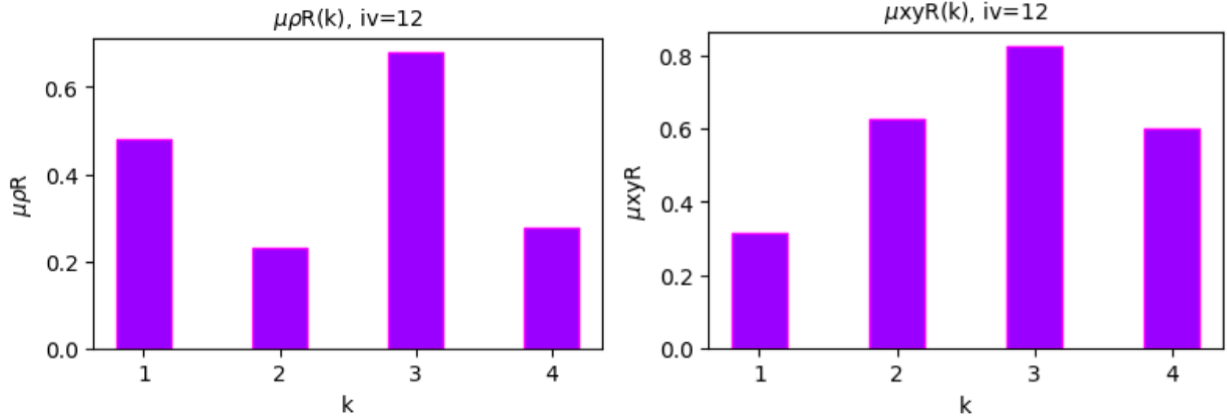

Fig. 8. The value of the degree of membership of a student with number $i_v = 12$ (Fig. 3) to clusters with numbers $k$, calculated on the basis of fuzzy membership functions $\mu_\rho(k, \rho 0)$ (Fig. 6) and $\mu_{xy}(k, dx0, dy0)$ (10)

For all clusters, the coordinates of the centers of clusters with numbers $k$ were also calculated based on the characteristics of "math score", "reading score" and the number of students $C_q$ in the clusters (Fig. 9).

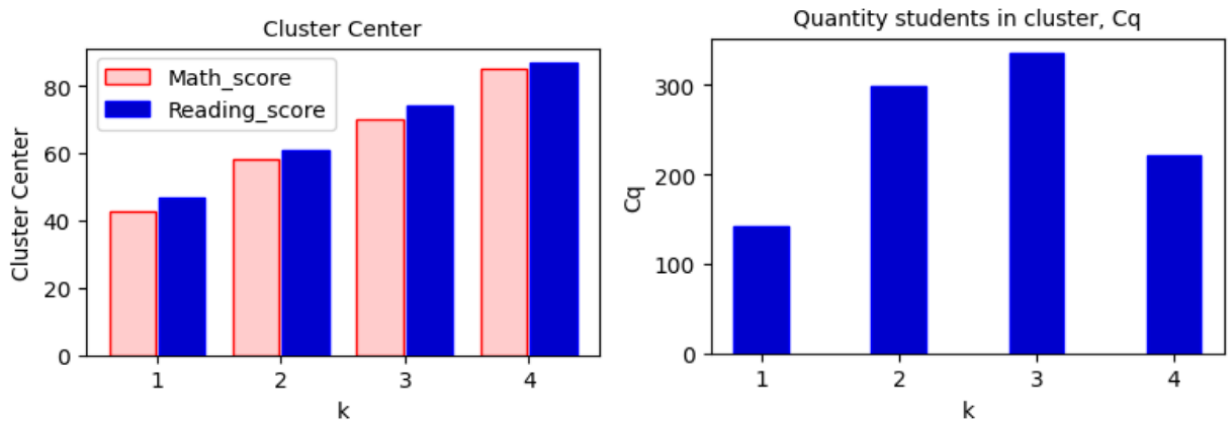

Fig. 9. Coordinates of the centers of clusters with numbers $k$ according to the characteristics "math score", "reading score" and the number of students $C_q$ in the clusters (Fig. 3)

The obtained coordinates of the cluster centers describe the average learning results of the students of each cluster. Analysis of the number of $C_q$ students in clusters allows to assess the level of difficulty of tasks. In the considered case, the distribution of the number of $C_q$ students in clusters is close to normal, so the difficulty level of the tasks is chosen correctly.

## 6. Conclusions

As a result of the work, the following results were obtained.

Software tools for clustering students according to their educational achievements using fuzzy logic have been developed. The program was implemented in Python using the Google Colab cloud service using the scikit-learn machine learning library and the scikit-fuzzy library for working with fuzzy logic. Clustering was performed in the space of two features using the classic K-Means method, which is characterized by simplicity and high speed.

The novelty of the work is the description of clusters by their own fuzzy triangular membership functions $\mu_\rho$, $\mu_x$, $\mu_y$ and $\mu_{xy}$ (5-10), which allowed to correctly determine the membership of each student in a certain cluster. The development of fuzzy membership functions of objects belonging to clusters is also done for educational purposes, as it allows a better understanding of the principles of using fuzzy logic. It is shown that the description of the shape of clusters using the radii $R_{cxL}$, $R_{cxR}$, $R_{cyDn}$, $R_{cyUp}$, calculated for various parameters, makes it possible to more accurately calculate the value of the fuzzy membership function of the object belonging to the cluster.

By the developed software, the test educational data was processed. As a result, students were divided into clusters, each of which corresponds to a group of students with a characteristic combination of grades in different subjects. Testing of the developed software showed that the use of fuzzy logic ensures the correct determination of the belonging degree of students to certain groups (clusters) even in the absence of clear boundaries between clusters. Due to this, the software allows you to individually determine the recommended level of difficulty of tasks for each student, depending on his previous learning results.

## Authors' Profiles

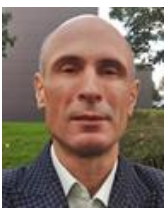

**Serhiy Balovsyak:** Graduated from Chernivtsi State University (1995). In 2018, he defended his doctoral dissertation in the specialty "Computer systems and components".

Currently position – associate professor at the Department of Computer Systems and Networks of Yuriy Fedkovych Chernivtsi National University, Ukraine.

Research Interests: digital processing of signals and images, programming, pattern recognition, artificial neural networks.

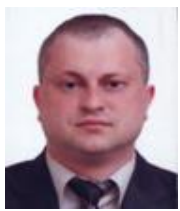

**Oleksandr Derevyanchuk:** Received the master of engineering degree (1999) and Ph.D. of Physics and Mathematics (2014) at the Yuriy Fedkovych Chernivtsi National University. He is a Candidate of Physical and Mathematical Sciences, Associate Professor of the Department of Professional and Technological Education and General Physics, Yuriy Fedkovych Chernivtsi National University, Chernivtsi, Ukraine.

Research Interests: digital processing of signals and images, programming, pattern recognition, artificial neural networks.

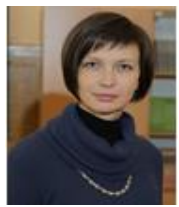

**Hanna Kravchenko:** PhD student at the of the Department of Professional and Technological Education and General Physics, Physical, Technical and Computer Sciences Institute of Yuriy Fedkovych Chernivtsi National University, Chernivtsi, Ukraine.

Research Interests: digital processing of signals and images, programming, pattern recognition, artificial neural networks.

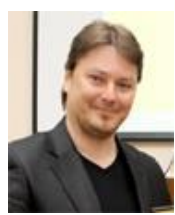

**Yuriy Ushenko:** M.Sc. in Telecommunications (2003). PhD in Optics and Laser Physics (2006). D.Sc. in Optics and Laser Physics, Taras Shevchenko National University of Kyiv (2015).

Current position – Professor, Head of Computer Science Department, Yuriy Fedkovych Chernivtsi National University, Ukraine.

Research Interests: Data Mining and Analysis, Computer Vision and Pattern Recognition, Optics & Photonics, Biophysics.

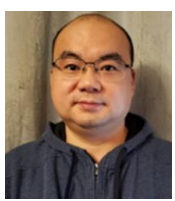

**Zhengbing Hu:** Prof., Deputy Director, International Center of Informatics and Computer Science, Faculty of Applied Mathematics, National Technical University of Ukraine "Kyiv Polytechnic Institute", Ukraine. Adjunct Professor, School of Computer Science, Hubei University of Technology, China. Visiting Prof., DSc Candidate in National Aviation University (Ukraine) from 2019. Major research interests: Computer Science and Technology Applications, Artificial Intelligence, Network Security, Communications, Data Processing, Cloud Computing, Education Technology.